\documentclass [a4paper,12pt,fleqn]{article}
\usepackage{graphicx}
\usepackage {subfigure,epsfig}

\usepackage {amsmath} \usepackage{amssymb} \usepackage{cite} \usepackage{amsthm}
\numberwithin{equation}{section} \numberwithin{table}{section} \mathindent=0pt
\theoremstyle{plain} \newtheorem{theorem}{Theorem}

\numberwithin{theorem}{section}

\begin{document}

\title{Exact solitary waves of \\
the Korteveg -- de Vries -- Burgers equation}
\author{Nikolai A. Kudryashov}
\date{Department of Applied Mathematics\\
Moscow  Engineering and Physics Institute\\
(State university)\\
31 Kashirskoe Shosse,  115409\\
Moscow, Russian Federation} \maketitle

\begin{abstract}
New approach is presented to search exact solutions of nonlinear
differential equations. This method is used to look for exact
solutions of the Korteveg -- de Vries -- Burgers equation. New
exact solitary waves of the Korteveg -- de Vries -- Burgers
equation are found.
\end{abstract}



\section{Introduction}

At the present great interest has been expressed by searching
exact solutions of nonlinear differential equations. This is
because many mathematical models are described by nonlinear
differential equations.

The inverse scattering transform \cite{1} and the direct method by
R. Hirota \cite{2} are known as impressive methods to look for
solutions of exactly solvable differential equations.

The singular manifold method \cite{3,4,5}, the transformation
method \cite{6,7} the tanh-function method \cite{8,9,10,11}, the
elliptic function method \cite{12,13}, and the Weierstrass
function method \cite{14,15} are useful in many applications to
search exact solutions of nonlinear partially solvable
differential equations.

In this letter we present new approach to search exact solutions
of nonlinear differential equations. The first advantage of our
approach that this one generalizes a number of known methods for
searching exact solutions of nonlinear ordinary differential
equations. The second advantage of the method is simplicity of
realization.

This letter has two parts. One of them is devoted to discussion of
our method. We apply our approach to look for new exact solutions
of the Korteveg -- de Vries -- Burgers equation in the second
part.

\section{Method applied}

First of all let us discuss our approach in detail. Consider
nonlinear ordinary differential equation in the polynomial form

\begin{equation}
M[y]=0\label{e:2.1}
\end{equation}

Assume we look for exact solutions of equation \eqref{e:2.1}. Take
nonlinear ordinary differential equation of lesser order than
equation \eqref{e:2.1}
\begin{equation}\label{e:2.2}
E[Y]=0
\end{equation}
We call any nonlinear ordinary differential equation \eqref{e:2.2}
of lesser order than \eqref{e:2.1} with known general solution as
the simplest equation.

First example of the simplest equation is the Riccati equation
\begin{equation}\label{e:2.3}
E[Y]=Y_z+Y^2- a Y-b=0
\end{equation}

We emphasize that the general solutions of equation \eqref{e:2.3}
have first order singularity. We are going to use this property of
equations \eqref{e:2.3} later.

 Suppose we have the relation between solutions of
 equations \eqref{e:2.1} and \eqref{e:2.2}
\begin{equation}\label{e:2.6}
y=F(Y)
\end{equation}
In addition, suppose that substitution \eqref{e:2.6} into equation
\eqref{e:2.1} leads to the relation in the form

\begin{equation}\label{e:2.7}
M[F(Y)]=\hat{A}E[Y]
\end{equation}
where $\hat{A}$ is a operator which we have to find.

From relation \eqref{e:2.7} we can see that for any solution of
the simplest equation \eqref{e:2.2} there exist a special solution
of equation \eqref{e:2.1} by the formula \eqref{e:2.6}.

To look for exact solutions of nonlinear ordinary differential
equation \eqref{e:2.1} in this letter we firstly introduce and use
formula \eqref{e:2.6} in the form

\begin{equation}
\begin{gathered}\label{e:2.14}
y(z)=A_0+A_1 Y+A_2 Y^2 + \ldots +A_n Y^n+\\
\\
+B_1 \left(\frac{Y_z}{Y}\right)+ B_2 \left(\frac{Y_z}{Y}\right)^2
+ \ldots +B_n \left(\frac{Y_z}{Y}\right)^n
\end{gathered}
\end{equation}
where $n$ is the singularity order of general solution of equation
\eqref{e:2.1} and $Y(z)$ is the general solution of equation
\eqref{e:2.3}, coefficients $A_k\,\,\,(k=0, \ldots, n)$ and
$B_k\,\,\,(k=1,\ldots, n)$ are found after substitution of
expression \eqref{e:2.14} into equation \eqref{e:2.1}. To find
coefficients $A_k\,\,\,(k=0, \ldots, n)$ and
$B_k\,\,\,(k=1,\ldots, n)$ we take the following simple theorem
into consideration.

\begin{theorem}
\label{T:2.1.} Let $Y(z)$ be solution of equation \eqref{e:2.3}
than equations

\begin{equation}
\label{2.18}Y_{zz} =2Y^3 - 3a Y^2 + (a^2 -2b) Y + ab
\end{equation}

\begin{equation}
\begin{gathered}
\label{e:2.19}Y_{{{\it zzz}}}=-6\, Y ^{4}+12\,
  Y ^{3}a+ \left( 8\,b-7\,{a}^{2}\right)
  Y  ^{2}+ \\
\left( -8\,ab+{a}^{ 3} \right) Y -2\,{b}^{2}+{a}^{2}b
\end{gathered}
\end{equation}

have special solutions that are expressed via the general solution
of equation \eqref{e:2.3}.
\end{theorem}

\begin{proof}Theorem 2.1 is proved by differentiation of \eqref{e:2.3} with respect to $z$
and substitution $Y_z$ from equation \eqref{e:2.3}  into
expressions obtained.

\end{proof}

Substituting expression \eqref{e:2.14} into equation \eqref{e:2.1}
and taking theorem 2.1 into account we have the relation

\begin{equation}
\begin{gathered}
\label{e:2.33}M[y]=\sum^{2\,n}_{k=0} P_k (a, b, A_0, \ldots,
A_n,B_1, \ldots, B_n)Y^{k-n}
\end{gathered}
\end{equation}

In the case of nontrivial solutions of the equations for
coefficients $A_k\,\,\,(k=0, \ldots, n)$ and
$B_k\,\,\,(k=1,\ldots, n)$

\begin{equation}
\begin{gathered}
\label{e:2.34}P_k (a, b, A_0, \ldots, A_n,B_1, \ldots,
B_n)=0,\,\,\, k=(0, \ldots, 2\,n)
\end{gathered}
\end{equation}
we have special solution of equation \eqref{e:2.1} by the formula
\eqref{e:2.14} where $Y(z)$ is general solution of equation
\eqref{e:2.3}.

In this manner the method applied allows us to look for exact
solutions of the origin equation \eqref{e:2.1}. Note that our
approach is the generalization of a number methods to look for
exact solutions of nonlinear differential equations. For example
using $B_k=0\,\,\,(k=1,\ldots,n)$ in \eqref{e:2.14} and $a=0$ in
\eqref{e:2.3} we have the tanh - function method as special case
of our approach.

\section{New exact solitary waves of \\
the Korteveg -- de Vries -- Burgers equation}

Let us apply our approach to look for exact solutions of the the
Korteveg -- de Vries -- Burgers equation. This equation takes the
form

\begin{equation}
\label{1.1}u_t +uu_{{x}}+\beta\,u_{xxx}=\alpha\,u_{{{\it xx}}}
\end{equation}

Nonlinear evolution equation \eqref{1.1} describes nonlinear waves
taking dispersion and dissipation into account. At $\nu=0$ and
$\beta\neq0$ we have the famous Korteveg -- de Vries equation from
equation \eqref{1.1}. The Causchy problem for this equation can be
solved by the inverse scattering transform \cite{1}. In the case
$\beta=0$ and $\nu\neq0$ we have the Burgers equation from
equation \eqref{1.1} that can be linearized by the Cole -- Hopf
transformation  into the heat equation \cite{16,17}. At
$\beta\neq0$ and $\nu\neq0$ equation \eqref{1.1} is not integrable
one because this one does not pass the Painleve test. However we
are going to look for exact solutions of this equation in this
letter.

There is the special solution in the form of the solitary wave of
equation \eqref{1.1} at $\beta\neq0$ and $\nu\neq0$ that was
firstly found in \cite{4}. Later this solution was rediscovered
time and again. However this exact solution has only one arbitrary
constant and this one is of limited usefulness.

Let us show that using our approach we can obtain more general
solitary waves.

Taking travelling wave into consideration

\begin{equation}
\label{1.1a}u(x, t) =y(z),\quad z=x-C_0 t
\end{equation}

we have from equation \eqref{1.1}

\begin{equation}
\label{1.2}C_{{1}}-C_{{0}}y-\alpha\,y_{{z}}+\beta\,y_{{{\it
zz}}}+\frac12\,{y}^{2}=0
\end{equation}

We can see that general solution of equation \eqref{1.2} has the
singularity order equal two and we have to take $n=2$ by formula
\eqref{e:2.14}. Therefore we will look for exact solution of
equation \eqref{1.2} in the form

\begin{equation}
\label{1.3}y \left( z \right) =A_{{0}}+A_{{1}}Y +A_{{2}} Y
^{2}+{\frac {B_{{1}}Y_{{z}}}{Y}}+{\frac {B
_{{2}}{Y_{{z}}}^{2}}{{Y}^{2}}}
\end{equation}

Taking equation \eqref{e:2.3} into account we have from equation
\eqref{1.2}

\begin{equation}
\begin{gathered}
\label{1.3a} y(z)=A_{{0}}+2\,B_{{1}}a+4\,B_{{2}}{a}^{2}-
2\,B_{{2}}b+\left (A_{{1}}-B_{{1}}-4\,B_{{2}}a\right
)Y+\\
\\
+\left (A_{{2}}+B_{{2}}\right )Y^{2}+{\frac
{B_{{1}}b+4\,B_{{2}}ab}{Y}}+{\frac {B_{{2}}{b}^{ 2}}{Y^{2}}}
\end{gathered}
\end{equation}

Substituting \eqref{1.3} into equation \eqref{1.2} we have

\begin{equation}
\begin{gathered}
\label{1.4a}B^{(1)}_2=-12\,\beta,\,\,\,B^{(2)}_2=0,
\end{gathered}
\end{equation}

Consider first case: $B_{{2}}=B^{(1)}_{{2}}=-12\,\beta$. We have

\begin{equation}
\begin{gathered}
\label{1.4b} B_{{1}}=-{\frac {12}{5}}\,\alpha+24\,\beta\,a
\end{gathered}
\end{equation}

\begin{equation}
\label{1.5}A_{{0}}=-4\,\beta\,{a}^{2}+C_{{0}}+\frac1{25}\,{\frac
{{\alpha}^{2}}{\beta}} +{\frac {12}{5}}\,\alpha\,a-16\,\beta\,b
\end{equation}

\begin{equation}
\label{1.6}A_{{1}}=\frac45\,{\frac
{\alpha\,{a}^{2}}{b}}-24\,\beta\,a-\frac85\,\alpha-{ \frac
{1}{125}}\,{\frac {{\alpha}^{3}}{b{\beta}^{2}}}
\end{equation}

\begin{equation}
\label{1.7}A_{{2}}^{(1)}=12\,\beta,\,\,\,A_{{2}}^{(2)}=0
\end{equation}

Assuming at the beginning $A_{{2}}=A_{{2}}^{(1)}=12\,\beta$ we
find additionally

\begin{equation}
\begin{gathered}
\label{1.8}b=-{\frac {1}{100}}\,{\frac
{100\,{\beta}^{2}{a}^{2}-{\alpha}^{2}}{{ \beta}^{2}}}
\end{gathered}
\end{equation}

\begin{equation}
\begin{gathered}
\label{1.9}C_{{1}}=\frac12\,{C_{{0}}}^{2}-{\frac
{18}{625}}\,{\frac {{\alpha}^{4}}{{\beta}^{2}}}
\end{gathered}
\end{equation}

As the result of calculations we obtain special solutions of the
Korteveg - de Vries - Burgers equation in the form

\begin{equation}
\begin{gathered}
\label{1.10}y \left( z \right) =C_{{0}}-12\,\beta\,{a}^{2}-
\frac{3}{125\,{\beta^{2}}}\left(\alpha-10\,\beta\,a\right)\left(\alpha+10\beta\,a\right)^{2}{Y}^{-1}+\\
\\
+ \frac{3}{25}{\frac
{\alpha^{2}}{\beta}}-\frac{12}{5}\alpha\,a-\frac{3}{2500\,\beta^{3}}
\left(\alpha-10\,\beta\,a\right)^{2}\left(\alpha+10\,\beta\,a\right)^{2}{Y}^{-2}
\end{gathered}
\end{equation}

where $Y(z)$ is determined by the formula

\begin{equation}
\begin{gathered}
\label{1.11}Y \left( z \right) =a + \frac {\alpha}{10\beta}\tanh
\left( \frac {\alpha}{10\beta} \left( z-C_{{2}} \right) \right)
\end{gathered}\end{equation}

and $Y(z)$ satisfies the equation

\begin{equation}
\label{1.12}Y_{{z}}+  Y^{2}-2\,aY +{\frac {1}{100}}\,{\frac
{100\,{\beta}^{2}{a}^{2}-{\alpha}^{ 2}}{{\beta}^{2}}}=0
\end{equation}

Exact solitary waves \eqref{1.11} are new solutions of equation
\eqref{1.2}. They have two arbitrary constants $a$ and $C_{{2}}$
and refer to the solitary waves in the form of kinks.

Assuming $B_{2}=B_{2}^{(1)}$ and $A_{2}=0$ we have relations

\begin{equation}
\label{1.13}A_{{1}}=-\frac {24\,\alpha}{5},\,\,\quad\,\,b=\frac
{\alpha^2}{400\,\beta^{2}},\,\,\quad\,\,a=0
\end{equation}

\begin{equation}
\label{1.16}C_{1}=\frac{1}{2}C_{{0}}^{2}-\frac
{18\,\alpha^4}{625\,\beta^{2}}
\end{equation}

\begin{equation}
\begin{gathered}
\label{1.17}y(z)=C_{{0}}+{ \frac {3}{50}}\,{\frac
{{\alpha}^{2}}{\beta}}-{\frac
{12}{5}}\,\alpha\,Y-12\,\beta\,Y^{2}-{\frac {3}{500}}\,{\frac {{
\alpha}^{3}}{{\beta}^{2}Y}}-{\frac {3}{40000}}\,{ \frac
{{\alpha}^{4}}{{\beta}^{3}Y^{2}}}
\end{gathered}
\end{equation}

Where $Y(z)$ is determined by the function

\begin{equation}
\label{1.17a}Y(z)= {\frac {\alpha}{{20}\,{\beta}}}\tanh \left(
\,{\frac {\alpha\,\left (z-C_{{2}}\right )}{20\,\beta}}\right)
\end{equation}

Substituting solution \eqref{1.17a} into \eqref{1.17} we have the
new  solution of the Korteveg - de Vries - Burgers equation again
but this solution is  the singular one.

\begin{equation}
\begin{gathered}
\label{1.18}y(z)=C_{{0}}+{\frac {3}{50}}\,{\frac
{{\alpha}^{2}}{\beta} }-{\frac {3}{25}}\,{\frac
{{\alpha}^{2}}{\beta}}\tanh \left(\,{\frac {\alpha\,\left
(z-C_{{2}}\right )}{20\,\beta}}\right)-\\
\\
-{\frac {3}{25}}\,{\frac{{ \alpha}^{2}}{\beta}}\left
(\tanh\left(\,{\frac {\alpha\,\left (z-C_{ {2}}\right
)}{20\beta}}\right)\right )^{-1}-{\frac {3}{100}}\,{\frac
{{\alpha}^{2}}{\beta}}\left (\tanh\left(\,{\frac {\alpha\,\left
(z-C_{{2}}\right )}{20\,\beta}}\right)
\right )^{2}-\\
\\
-{\frac {3}{100}}\,{\frac {{\alpha}^{2}}{\beta}}\left
(\tanh\left(\,{\frac {\alpha\,\left (z-C_{{2}}\right
)}{20\,\beta}}\right)\right )^{-2}
\end{gathered}
\end{equation}

Consider case: $B_{{2}}=0$. Now we have to obtain the known
solution. After calculations we find

\begin{equation}
\begin{gathered}
\label{1.19}B_{{1}}=0,\,\,\quad \,
A_{{2}}=-12\,\beta,\,\,\quad\,\,A_{{1}}=24\,a\,\beta-\frac{12}{5}\,\alpha
\end{gathered}
\end{equation}

\begin{equation}
\begin{gathered}
\label{1.21}A_{{0}}=C_{{0}}+{\frac
{12}{5}}\,\alpha\,a-4\,\beta\,{a}^{2}+\frac{1}{25}\,{ \frac
{{\alpha}^{2}}{\beta}}+8\,\beta\,b
\end{gathered}
\end{equation}

\begin{equation}
\begin{gathered}
\label{1.22}b=-{\frac {1}{100}}\,{\frac
{100\,{\beta}^{2}{a}^{2}-{\alpha}^{2}}{{ \beta}^{2}}}
\end{gathered}
\end{equation}

\begin{equation}
\label{1.23}C_{1}=\frac{1}{2}C_{{0}}^{2}-\frac
{18\,\alpha^4}{625\,\beta^{2}}
\end{equation}

Exact solution of the Korteveg - de Vries - Burgers equation is
found by the formula

\begin{equation}
\begin{gathered}
\label{1.24}y(z)=C_{{0}}+{\frac {12}{5}}\,\alpha\,a-12
\,\beta\,{a}^{2}+{\frac {3}{25}}\,{\frac
{{\alpha}^{2}}{\beta}}+\left (24\,\beta\,a-{\frac {12}{5}}\,
\alpha\right )Y-12\,\beta\,Y^{2}
\end{gathered}
\end{equation}

Where $Y(z)$ is determined by the function

\begin{equation}
\begin{gathered}
\label{1.25}Y(z)=a \frac{1}{10}\, {{\frac
{{\alpha}}{{\beta}}}}\tanh\left( \frac{1}{10}\, {{\frac
{{\alpha}}{{\beta}}}}\left (z-C_{{2}}\right )\right)
\end{gathered}
\end{equation}

It is hoped that we have two arbitrary constants again but this is
not the case because substituting solutions \eqref{1.25} into
\eqref{1.24} we obtain known solution with one arbitrary constant
in the form

\begin{equation}
\begin{gathered}
\label{1.26}y(z)=C_{{0}}+{\frac {3}{25}}\,{\frac
{{\alpha}^{2}}{\beta}}-{\frac {3}
{25}}\,\frac{{\alpha}^{2}}{\beta}\left (\tanh\left( {{\frac
{{\alpha}}{{10 \beta}}}}\left
(z-C_{{2}}\right )\right)\right )^{2}+\\
\\
+ {\frac {6}{25}}\,{{\frac {{\alpha}^{2}}{{\beta}}}}\tanh\left( \
{\frac {\alpha}{{10\,\beta}}}\left (z-C_{{2}}\right )\right)
\end{gathered}
\end{equation}

The last solution is solitary wave in the form of kink with one
arbitrary constant at given $\beta$ and $\nu$. This solution was
found in work \cite{4} and rediscovered more than once later.

\section{Conclusion}

Let us emphasize in brief the results of this work. In this paper
we presented new approach to look for exact solutions of nonlinear
ordinary differential equations that we called the simplest
equation method. The idea of our approach is to use simplest
nonlinear equation with known general solution in order to express
special solution of nonlinear differential equation of higher
order. This method was applied to look for exact solitary waves of
the Korteveg - de Vries - Burgers equation. Application of our
method allowed us to find new exact solitary waves with two
arbitrary constants of the Korteveg - de Vries - Burgers equation.

\section {Acknowledgments}

This work was supported by the International Science and
Technology Center under Project No 1379-2.

\end{document}